\documentstyle[preprint,aps,psfig]{revtex}
\begin{document}
\tolerance=10000
\draft
\title{Simulation of a semiflexible polymer 
in a narrow cylindrical pore}
\author{Dominique J. Bicout}
\address{Theory Group, Institut
Laue-Langevin, BP 156, F-38042 Grenoble Cedex 9, France}
\author{Theodore W. Burkhardt} 
\address{Department of Physics, Temple University,
Philadelphia, PA 19122, U. S. A.}
\maketitle
\begin{abstract}
The probability that a randomly accelerated particle in two dimensions 
has not yet left a simply connected domain ${\cal A}$ after a time $t$ 
decays as $e^{-E_0t}$ for long times. The same quantity $E_0$ also 
determines the confinement free energy per unit length
$\Delta f=k_BT\thinspace E_0$ of a semiflexible polymer in a narrow 
cylindrical pore with cross section ${\cal A}$. From simulations of 
a randomly accelerated particle we estimate the universal
amplitude of $\Delta f$
for both circular and rectangular cross sections.  
\end{abstract}
\vskip 0.75cm
\centerline{PACS 36.20.Ey, 05.10.-a, 05.10.Gg, 05.70.Ce}
\vskip 0.5cm
\centerline{Submitted to J.Phys. A}\eject

\narrowtext
Consider a long semiflexible polymer with persistence length $P$
fluctuating
in a cylindrical pore with diameter $D$. In the narrow-pore limit $D<<P$
the free energy of confinement per unit length $\Delta f$ is given by
\begin{equation}
\Delta f=A_{\bigcirc}\ {k_BT\over P^{1/3}D^{2/3}}.\label{e1}
\end{equation}
This follows from simple scaling or dimensional arguments
\cite{o,dfl,b}, such as that given below.
Similarly, for a pore with a rectangular cross section \cite{b} with
edges $L_1,L_2<<P$
\begin{equation} \Delta
f=A_{\framebox[2mm]{}}\ {k_BT\over
P^{1/3}}\left(\thinspace{1\over L_1^{2/3}}+
{1\over L_2^{2/3}}\thinspace\right).\label{e2}
\end{equation}

The dimensionless constants $A_{\bigcirc}$, $A_{\framebox[2mm]{}}$ in
equations (\ref{e1}) and (\ref{e2}) are
universal numbers, independent of both macroscopic and microscopic
properties of the polymer chain. From computer simulations Dijkstra et
al. estimated \cite{dfl}
\begin{equation}
A_{\bigcirc}=2.46\pm 0.07.\label{e3}
\end{equation}
Solving an integral equation numerically
that arises in an exact analytical approach,
Burkhardt obtained \cite{b}
\begin{equation}
A_{\framebox[2mm]{}}=1.1036.\label{e4}
\end{equation}

In the Monte Carlo simulations of Dijkstra et al. \cite{dfl},
polymer configurations consistent with the Boltzmann distribution
were generated with a Metropolis algorithm incorporating detailed
balance. In this Letter we estimate
$A_{\bigcirc}$, $A_{\framebox[2mm]{}}$ with a precision of about $1\%$ 
using a Langevin dynamics approach, which is simple and efficient.
Instead of a confined semiflexible polymer, we simulate a
Newtonian particle which is randomly accelerated by Gaussian white noise
in
two dimensions. In the narrow-pore or tight-confinement limit
these two systems have equivalent statistical properties, as
discussed in \cite{b} and reviewed below. The basic idea, illustrated in
figure 1, is that each of the possible world lines of a particle, which
is randomly accelerated in two dimensions and remains in a
domain ${\cal A}$ for a time $t$, corresponds
to an allowed configuration of a tightly-confined semiflexible polymer
in a cylindrical pore or tube with cross section ${\cal A}$.

We now describe the correspondence in more detail.
As in \cite{b}, polymer configurations are specified in terms
of Cartesian coordinates $({\bf x},t)=(x_1,x_2,t)$. The $t$ axis is 
parallel to the axis of the cylindrical pore, as shown in figure 1.
In the narrow-pore limit
$D<<P$ or $L_1,L_2<<P$, configurations
with overhangs are negligible, i.e.
${\bf x}$ is a single-valued function of $t$. The partition
function is given by the path integral \cite{b}
\begin{equation}
Z({\bf x},{\bf u};{\bf x_0},{\bf u_0};t)=\int_{\cal A} D^2x\thinspace
\exp\Big[-{1\over 2}P\int_0^t dt\thinspace
\left({d^2{\bf x}\over dt^2}\right)^2\Big]\label{e5}
\end{equation}
where ${\bf x}$ and ${\bf u}=d{\bf x}/dt$ denote the displacement
and slope of the polymer at $t$, and ${\bf x_0}$ and ${\bf u_0}$
the same quantities at $t=0$. A hard-wall confining potential is
assumed, and the ${\bf x}$ integration is limited to the domain ${\cal
A}$.

The path integral implies the
the partial differential equation \cite{b,gb,fh,f,mhl,r}
\begin{equation}
\Big[{\partial\over\partial t}+{\bf u\cdot\nabla_x}
-{1\over 2P}{\bf\nabla_u}^2\Big]\thinspace Z({\bf x},{\bf u};{\bf x_0}
{\bf u_0};t)=0\label{e6}
\end{equation}
to be solved with the initial condition
\begin{equation}
Z({\bf x},{\bf u};{\bf x_0},{\bf u_0};0)=
\delta({\bf x}-{\bf x_0})\delta({\bf u}-{\bf u_0}).\label{e7}
\end{equation}
Since the polymer is confined to the interior of the pore and
configurations with a  discontinuity in slope cost an infinite energy,
$Z({\bf x},{\bf u};{\bf x_0},{\bf u_0};t)$
vanishes for ${\bf u\cdot n}>0$ as ${\bf x}$ approaches a hard wall.
Here ${\bf n}$ is a vector normal to the wall and directed toward
the interior  of the pore.

As in \cite{b} we consider exponentially decaying solutions of equation
(\ref{e6}) with the form $\psi({\bf x},{\bf u})\exp(-Et)$. The
eigenfunctions $\psi$ and eigenvalues $E_n$ satisfy
\begin{equation}
\Big[{\bf u\cdot\nabla_x}-{1\over 2P}
{\bf\nabla_u}^2-E_n\Big]\psi_n({\bf x},{\bf u})=0\label{e8}
\end{equation}
where $\psi({\bf x},{\bf u})$ also vanishes for ${\bf u\cdot n}>0$
as ${\bf x}$ approaches the pore wall.
In the long-polymer limit the partition function
and the confinement free energy per unit length are given by
\begin{equation}
Z({\bf x},{\bf u};{\bf x_0},{\bf u_0};t)\approx
{\rm const}\times\psi_0({\bf x},{\bf u})
\psi_0({\bf x_0},{\bf -u_0})e^{-E_0 t}\quad\quad t\to\infty\label{e9}
\end{equation}
\begin{equation} {\Delta f\over
k_BT}=E_0(P,D)\label{e10}
\end{equation}
for a pore with a circular cross section of diameter $D$, where $E_0$ is
the smallest of the eigenvalues $E_n$.

The $P$ and $D$ dependence in equation (\ref{e1}) may be derived from
(\ref{e10}) by making the scale change ${\bf x}=
\alpha{\bf x'}$, ${\bf u}=\beta{\bf u'}$ in (\ref{e8}),
where $\alpha$ and $\beta$ are arbitrary positive constants. This leads
to the scaling property
\begin{equation}
E_n(P,D)=\alpha^{-1}\beta E_n(\alpha^{-1}\beta^3
P,\alpha^{-1}D)\label{e11}
\end{equation}
of the eigenvaues.
Setting $\alpha =D$ and $\beta =P^{-1/3}D^{1/3}$ in equation (\ref{e11})
and substituting the result in (\ref{e10}), we obtain
\begin{equation} {\Delta f\over k_BT}={E_0(1,1)
\over P^{1/3}D^{2/3}}\label{e12}
\end{equation}
This expression is entirely consistent with equation (\ref{e1}) and
implies
\begin{equation}
A_{\bigcirc}=E_0(1,1).\label{e13}
\end{equation}
Similarly, for a pore with a rectangular cross section
\begin{equation}
2A_{\framebox[2mm]{}}=E_0(1,1,1)\label{e14}
\end{equation}
where $E_0(P,L_1,L_2)$ denotes the smallest eigenvalue in equation
(\ref{e8})
for a rectangular domain with edges $L_1$, $L_2$.
Equations (\ref{e13}) and (\ref{e14}) allow us to determine the
universal amplitudes $A_{\bigcirc}$ and $A_{\framebox[2mm]{}}$ from
calculations with $P=D=L_1=L_2=1$.
   
For the rectangular domain equation (\ref{e8}) has
separable eigenfunctions $\psi_{m,n}({\bf x},{\bf u})=
\phi_m(x_1,u_1)\phi_n(x_2,u_2)$, with eigenvalues 
$E_{m,n}=E_m^{({\rm 1\thinspace dim})}(P,L_1)+
E_n^{({\rm 1\thinspace dim})}(P,L_2)$, where 
\begin{equation}\Big[u{\partial\over\partial x}-{1\over 2P}
{\partial^2\over\partial u^2}-E_m^{({\rm 1\thinspace dim})}(P,L)
\Big]\phi_m(x,u)=0.\label{e14b}
\end{equation}
Here $0<x<L$, and $\phi(x,u)$ vanishes 
for $x=0,u>0$ and $x=L,u<0$. Thus equation (\ref{e14}) may be rewritten
as
\begin{equation}
A_{\framebox[2mm]{}}=E_0^{({\rm 1\thinspace dim})}(1,1)\label{e14c}
\end{equation}
in terms of the smallest eigenvalue of (\ref{e14b}) for $P=L=1$.
The earlier numerical result for $A_{\framebox[2mm]{}}$, 
noted in equation (\ref{e4}), 
was obtained in \cite{b} by converting (\ref{e14b}) to
an integral equation, determining the smallest eigenvalue numerically,
and 
substituting the result in (\ref{e14c}).   

Now consider a particle which is randomly accelerated in the
$d$-dimensional 
space $(x_1,\dots x_d)$ by Gaussian white noise with zero mean
according to
\begin{equation}
{d^2 x_i\over dt^2}=\xi_i(t)\quad\quad
\langle\xi_i
(t)\xi_j(t')\rangle=P^{-1}\delta_{ij}\delta(t-t').\label{e14d} 
\end{equation}
The probability density 
in phase space $({\bf x},{\bf u})$ that the particle
remains in the domain ${\cal A}$ for a
time $t$ while the position and velocity evolve
from $({\bf x_0},{\bf u_0})$ to  $({\bf x},{\bf u})$ satisfies a
Fokker-Planck equation \cite{r} with exactly the same form (\ref{e6}),
initial condition (\ref{e7}), and boundary condition at the boundary of
${\cal A}$. Thus the probability density of the randomly accelerated
particle equals the partition function of a semiflexible polymer in 
a pore with cross section 
${\cal A}$ and decays as $\exp(-E_0t)$ for long times, 
as in equation (\ref{e9}).

To estimate $A_{\framebox[2mm]{}}$ using equation (\ref{e14c}), 
we simulated a randomly accelerated particle in one spatial 
dimension. From simulations in two dimensions we obtained a second
estimate of
$A_{\framebox[2mm]{}}$ based on (\ref{e14}) and a prediction for 
$A_{\bigcirc}$ from (\ref{e13}). The simulation routine, similar
to that in \cite{bb}, will now be described briefly. 
 
In an unbounded $d$-dimensional space the exact solution of
the Fokker-Planck equation (\ref{e6}) with initial condition (\ref{e7}) 
and with $P=1$ is given by \cite{gb,m}
\begin{equation}
Z_{\rm free}({\bf x},{\bf u};{\bf x_0},{\bf u_0};t)=
\left({3^{1/2}\over\pi t^2}\right)^d\,
\exp\left\{-\frac{6}{t^3}\,\left[({\bf x}-{\bf x_0}-{\bf u_0}t)\cdot
({\bf x}-{\bf x_0}-{\bf u}t)+
\frac{1}{3}({\bf u}-{\bf u_0})^2t^2\right]\right\}\label{e15}
\end{equation}
Trajectories with the probability distribution
$Z_{\rm free}({\bf x}_{n+1},{\bf u}_{n+1};{\bf x}_n,{\bf
u}_n;\Delta_{n+1})$
given by (\ref{e15}) are generated using the algorithm
\begin{eqnarray}
{\bf x}_{n+1}&=&{\bf x}_n+{\bf u}_n\Delta_{n+1}+
{\bf e}_n\thinspace\Delta_{n+1}^{3/2}\,
\left( 3^{-1/2}s_{n+1}+r_{n+1}\right)\label{e16} \\
{\bf u}_{n+1}&=&{\bf u}_n+{\bf e}_n\thinspace
2\Delta_{n+1}^{1/2}\,r_{n+1}\label{e17}
\end{eqnarray}
where ${\bf x}_n$ and ${\bf u}_n$ are the position
and velocity of the particle at time $t_n$, and
$\Delta_{n+1}=t_{n+1}-t_n$. The quantity ${\bf e}_n$ is a
unit vector that points either along the positive $x_1$ axis , the
positive $x_2$
axis, $\dots$, or the positive $x_d$ axis
with equal probability, and
$r_n$ and $s_n$ are independent Gaussian random
numbers satisfying
\begin{equation}
\langle r_n\rangle=\langle s_n\rangle=0,\quad
\langle r_n^2\rangle=\langle s_n^2\rangle=1.\label{e18}
\end{equation}

In the absence of boundaries there is no time-step error in the above
algorithm,i.e., the $\Delta_n$ may be chosen arbitrarily. Close to 
the boundaries small time steps are needed.
As in \cite{bb} we performed our simulations with
\begin{equation}
\Delta_{n+1}=10^{-5}+10^{-1}D_n\label{e19}
\end{equation}
where $D_n$ is the distance from the particle to the closest point
of the domain boundary at time $t_n$.
This time step fulfills
a reliability criterion discussed in
\cite{bb}.

Some sample simulational results are shown in figure 2.
The quantity $Q(t)$ is the probability that a particle
with a random initial position in a one or two-dimensional domain and 
with initial velocity zero, 
which is randomly accelerated
according to equation (\ref{e14d}) with $P=1$, has not yet left the 
domain after a time $t$. The curves labelled 'circle,' 'square,' and 
'interval' refer to a circular domain of diameter
$D=1$, a square with edges $L_1=L_2=1$, and a one-dimensional interval
of
length $L=1$. Each of curves is based on 10,000 independent
trajectories.
 
Since $Q(t)=\int d^dx\int d^du\int d^dx_0\thinspace
Z({\bf x},{\bf u};{\bf x_0},{\bf 0};t)$, it decays 
as $e^{-E_0t}$ for long times, with the same decay constant $E_0$ 
as $Z({\bf x},{\bf u};{\bf x_0},{\bf u_0};t)$ in equation (\ref{e9}).
We determined $E_0$ for the circular, square, and one-dimensional
domains
by fitting $Q(t)$ for long times with an exponential function.
A surprising result, shown in figure 2, is that the curves for the 
circular and square domains practically coincide when plotted 
versus $E_0t$ instead of $t$. 

We also estimated $E_0$ for circular, square, and one-dimensional 
domains from trajectories that all begin at the the center of the 
domain with initial velocity zero. The results are consistent with the 
results for random initial positions but have a somewhat greater
statistical 
uncertainty.    

Combining our best estimates of $E_0$ with equations (\ref{e13}),
(\ref{e14}), 
and (\ref{e14c}), we obtain 
\begin{eqnarray}
A_{\bigcirc}&=&2.375\,\pm\,0.013\label{e20}\\
A_{\framebox[2mm]{}}&=&1.108\,\pm\,0.013\label{e21}
\end{eqnarray}

The result (\ref{e20}) for $A_{\bigcirc}$ is somewhat lower than
the estimate (\ref{e3}) of Dijkstra et al. \cite{dfl} and has a smaller
statistical uncertainty.
The result (\ref{e21}) for $A_{\framebox[2mm]{}}$ in equation is
consistent
with the value (\ref{e4}) obtained by Burkhardt from the numerical
solution
of an exact integral equation \cite{b}. Bundschuh \cite{rb} has
also confirmed (\ref{e4}) to within a few percent with a numerical
transfer matrix approach \cite{bll} for semiflexible polymers.

In summary we have found a simple and efficient simulational procedure
for calculating the confinement free energy of a semiflexible
polymer in a narrow cylindrical pore with cross section ${\cal A}$.
We use the equivalent statistical properties of a Newtonian particle
which is
randomly accelerated by Gaussian white noise in two dimensions. The
probability
that the particle has not yet left a domain ${\cal A}$ in a time $t$
decays
as $e^{-E_0t}$. We determine $E_0$ from our simulations and then
interpret
it, following equation (\ref{e10}), in terms of the polymer free energy.
We emphasize that the equivalence between the statistics of the polymer
and the randomly accelerated particle is asymptotically exact in the
limit in which the pore
diameter is much smaller than the polymer persistence length.

TWB appreciates the hospitality of the Institut Laue-Langevin,
where this work was begun, and thanks Ralf Bundschuh for helpful
correspondence.

\begin{figure}[ht]
\vspace{0.3cm}
\centerline{\psfig{figure=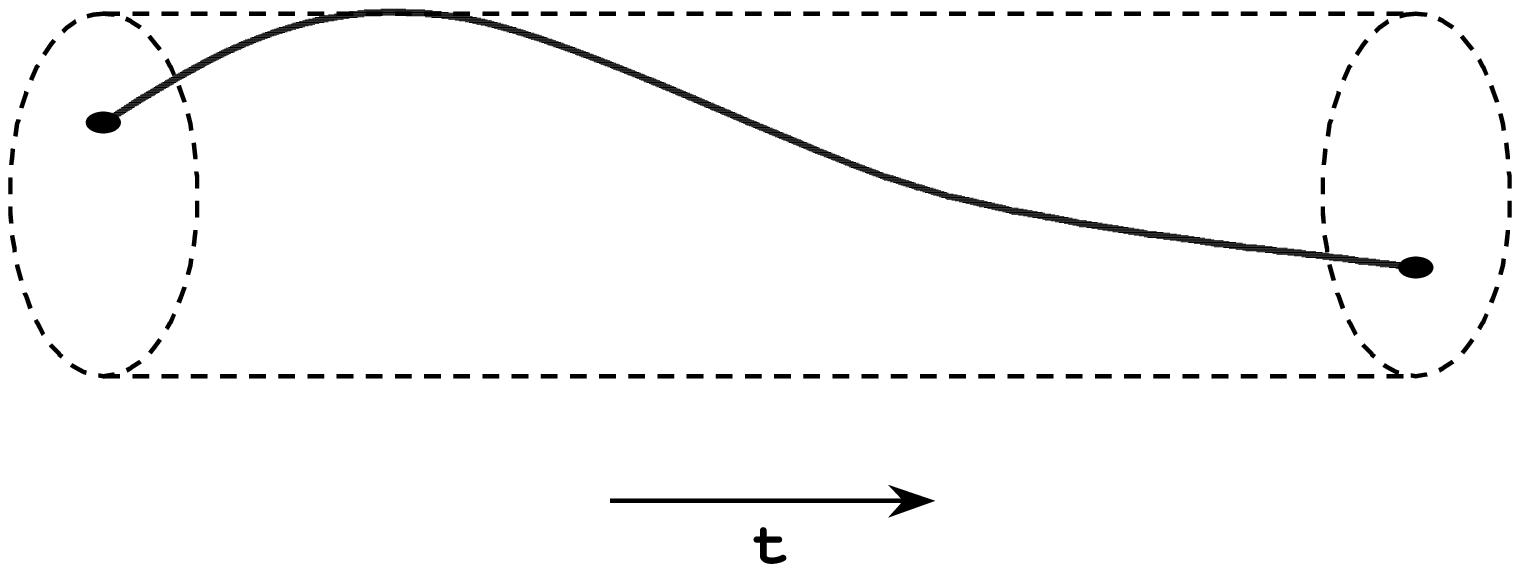,width=3.8in,angle=0}}
\vspace{-5cm}
\caption{The curve may be interpreted as a tightly-confined semiflexible 
polymer in a cylindrical pore with cross section ${\cal A}$ or as
the world line of a particle which is randomly accelerated in two
dimensions and remains
in a domain ${\cal A}$ for a time $t$.}
\label{fig1}
\end{figure}
\vspace{1.cm}

\newpage

\begin{figure}[ht]
\vspace{0.3cm}
\centerline{\psfig{figure=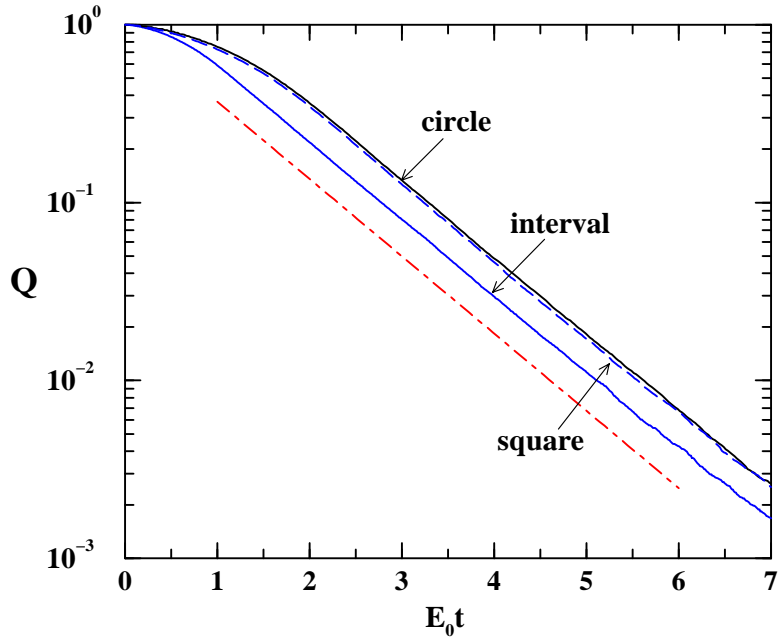,width=4.0in,angle=0}}
\vspace{0.25cm}
\caption{Probability $Q(t)$ that a particle 
with random initial position in a one or two-dimensional
domain and initial velocity zero, which is randomly accelerated
according to equation (17) with $P=1$,
has not yet left the domain after a time $t$.
The curves labelled 'circle' (solid line), 'square' (dashed line), 
and 'interval' (solid line) correspond to a circular domain with
diameter 1, a
square domain with edge 1, and a one-dimensional interval with length 1. 
For the three curves 
$E_0=2.375,\ 2.199,\ 1.108$, respectively. The dot-dashed line
represents a pure 
exponential decay $e^{-E_0t}$.} 
\label{fig2}
\end{figure}

\end{document}